\documentclass[review]{elsarticle}

\usepackage[toc,page]{appendix}
\usepackage{lineno,hyperref}
\usepackage{amsmath}
\usepackage{longtable}
\usepackage{supertabular}
\usepackage{graphicx} 
\modulolinenumbers[5]
\usepackage{setspace}
\usepackage{multirow,booktabs}

\journal{Acta Materialia}

\begin{document}
\begin{frontmatter}

\title{Eutectic Growth in Two-Phase Multicomponent Alloys}
\author[mainaddress]{Oriane Senninger \corref{correspondingauthor}}
\address[mainaddress]{Department of Materials Science and Engineering, Northwestern University, 2220 Campus Drive, Evanston, IL 60208, USA}
\cortext[correspondingauthor]{Corresponding author}
\ead{oriane.senninger@insp.jussieu.fr}

\author[mainaddress]{Peter W Voorhees}


\begin{abstract}
A  theory of two-phase eutectic growth for a multicomponent alloy is presented. This theory employs the thermodynamic equilibrium at the solid/liquid interface and thus makes it possible to use standard CALPHAD databases to determine the effects of multicomponent phase equilibrium on eutectic growth. Using the same hypotheses as the Jackson Hunt theory, we find that the growth law determined for binary alloys in the Jackson Hunt theory can be generalized to systems with N elements. In particular, a new model is derived from this theory for ternary two-phase eutectics. The use of this model to predict the eutectic microstructure of systems is discussed.  
\end{abstract}

\end{frontmatter}

\section{Introduction}

Eutectic alloys possess many advantages compared to single phase systems. Indeed, they have a low melting point compared to pure components and their composite microstructure procure them superior mechanical properties.

For binary eutectics,  Hillert \cite{Hillert1957} and later Jackson and Hunt \cite{Jackson1966} determined a scaling parameter of the microstructure at a given solidification velocity. Moreover, they established the link between this parameter and thermodynamic and thermophysical properties of alloys.  
This scaling parameter has been proved to be relevant to characterize the eutectic microstructure of many regular binary alloys \cite{Magnin1991}.  

However, a analogous theory for alloys with many components and growing as a two-phase eutectic does not exist. Such multicomponent two-phase eutectics are common and have been studied in, Al-Cu-Ag \cite{DeWilde2004}, Fe-Si-Mn, Fe-Si-Co \cite{Yamauchi1996}, Al-Cu-Ni \cite{Rinaldi1972} and Ni-Al-Cr-Mo \cite{Raj2001}. Moreover, most commercially relevant materials contain still more alloying elements. Unfortunately, a comprehensive model for the growth of these multicomponent two-phase eutectics does not exist.  
  However, there has been progress towards a general theory. Catalina et al. \cite{Catalina2015} proposed a model for  eutectic growth of two-phase eutectics containing N elements, but restricted the treatment to the case where one of the phases has no solid solubility for the solute elements. 
  Fridberg and Hillert \cite{Fridberg1970} published a model for the growth process of a binary alloy containing a small amount of an additional element.
 In  ternary alloys, McCartney et al. \cite{McCartney1980} and DeWilde et al. \cite{DeWilde2005} have given two different models. In the McCartney-Hunt model,  simplifying approximations were employed on the alloy phase diagram  and  the diffusion process.   DeWilde et al.  employed an approximation for the manner in which the long-range diffusion field decays and for  concentration profiles in the liquid phase.  While all of these treatments provide important insights into eutectic solidification of multicomponent alloys, they lack the generality needed for many applications.  
  
In this paper, we present a  method to compute the mean undercooling of a two-phase eutectic as a function of the eutectic spacing and the velocity for any alloy containing N elements in the spirit of the Jackson Hunt model (Section II). This general method removes the approximations introduced in the models \cite{Catalina2015} \cite{McCartney1980} \cite{DeWilde2005} mentioned above. It is then applied to binary alloys and compared to the Jackson Hunt theory in section III. The model derived from this general method for ternary alloys is given in section IV. We finally discuss in section VI the use of this model as a way to predict of the eutectic microstructure evolution of an alloy with the addition of a new element. We conclude this paper by a summary of results presented and possible future continuation of this work.

\section{Two-phase eutectic growth of alloys with N elements \label{sec:Theory}}

In this section we present our general methodology to compute the mean undercooling of any two-phase eutectic alloy with N elements. 

\begin{figure}[h!]
\centering
\includegraphics[width=0.7\textwidth]{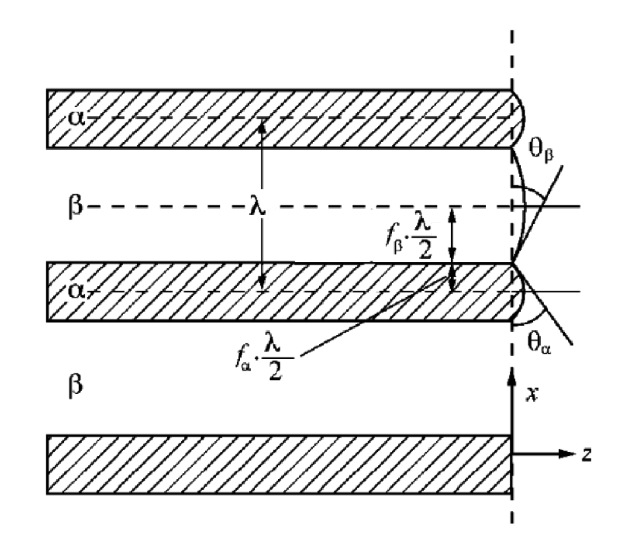}
\caption{\label{fig:intro} Schematic representation of steady state directional growth with a lamellar morphology. Quantities reported on the figure are: eutectic spacing $\lambda$, solid fraction of $\alpha$ phase ($f_\alpha$) and $\beta$ phase ($f_\beta$), angles of curvature of $\alpha$ phase ($\theta_\alpha$) and $\beta$ phase ($\theta_\beta$) at the tri-junction. (After Ludwig et al. \cite{Ludwig2004})}
\end{figure}

We study the directional solidification at steady state of a two-phase eutectic with an initial concentration $\left(C^\infty_2,\dots,C^\infty_N \right) $. We assume that this eutectic develops a lamellar morphology such as the one presented in Fig \ref{fig:intro}. \\

Here the eutectic temperature ($T_E$) is defined as the thermodynamic equilibrium temperature of the solid-liquid interface at steady state, which depends on the alloy initial composition. All quantities referring to this temperature will be identified with a 'E' superscript.   
We assume that for any position $x$ at the interface, the solid/liquid interface is at thermodynamic equilibrium at a temperature $T_u(x)$. 
So for any position $x$ of the interface, the chemical potentials of any specie $i=1..N$ in the liquid phase and in the solid phase $\phi$ are equal: 
\begin{equation}
\mu^\phi_i(C^\phi_2,..,C^\phi_N,T_u,p^\phi)=\mu^l_i(C^l_2,..,C^l_N,T_u,p^l) \;\;i=1,2\ldots N
\label{slequilibrium}
\end{equation}
where $C_i$ is the mole fraction of component $i$, $p$ is the pressure, and $\phi$ can be either one of the two solid phases. For a given phase,  assuming that $C^l_2..,C^l_N, p^l$ and $p^\phi$ are known,  this gives $N$ equations and $N$ unknowns. Thus once the composition of the liquid at the interface and the pressure in the solid phase are known, and by assuming that $p^l $ does not change from that at the equilibrium state, the composition of the solid phase is known and the undercooling is fixed. 

The variations in the rejection of solutes in front of solid phases, $\alpha$ and $\beta$,  induce changes in the concentrations in the liquid phase at the interface compared with the equilibrium state, $(C^{lE}_2,..,C^{lE}_N)$. In addition, the interface curvature due to the surface energies equilibrium at the trijunctions (points where the two solid phases are in contact with the liquid phase) induces a variation of the internal pressure in solid phases. Since local equilibrium is assumed to hold,  these variations in the liquid composition induce changes in concentrations in solid phases from their equilibrium values, and a change in  the interface temperature from $T_E$.  The compositions of the solid, liquid and the temperature are related by  $N$ chemical potential equations for each solid phase.  Unfortunately, these equations are nonlinear, and thus we assume small deviations from the equilibrium temperature, and phase compositions to relate the solid phase compositions and undercooling temperature to the liquid composition.  
The development of these N 
 equalities (\ref{slequilibrium})  for each phase is given in the appendix \ref{Appendix A}. 
 This development leads to a matrix  expressing the change in the concentration in solid phases from equilibrium, $\Delta C^\phi_i=C^{\phi E}_i-C^\phi_i$ ($i=2 \dots N$) and the undercooling $\Delta T=T_E-T_u$ as a function of  the concentration in the liquid phase $\Delta C^l_i=C^{l E}_i-C^l_i$ and of pressure in the solid phase $\Delta p^\phi$.
 
At a given point $x$ along the interface, the undercooling $\Delta T$ is thus expressed as a sum of a solutal ($ \Delta T_C$) and a curvature ($ \Delta T_R$) undercooling (see appendix \ref{Appendix A}) :
\begin{equation}
\Delta T(x)=\Delta T_C(x)+\Delta T_R(x)
\label{Tgeneral}
 \end{equation}
where
\begin{eqnarray}
\Delta T_C(x)&=& \displaystyle \sum^N_{i=2} m^\phi_i(C^{lE}_i-C^l_i(x)) \label{dTsolute}\\
\Delta T_R(x)&=& -\frac{V^\phi_m}{\Delta S_{\phi l}}\Delta p^\phi(x) \label{dTcurvature}
\end{eqnarray}
where $m^\phi_i$ is a slope of a liquidus surface, $V^\phi_m$ is a molar volume,  and $\Delta S_{\phi l}$  are defined in appendix \ref{Appendix A} as functions of derivatives of molar Gibbs free energies of the solid and liquid phases.
As $\Delta p^\phi=-\sigma_{\phi l} \kappa (x)$ where $\sigma_{\phi l}$ is the $\phi/l$ surface energy and $\kappa (x)$ is the interface curvature at $x$, Eq. (\ref{dTcurvature}) can be  re-written:
\begin{equation}
\Delta T_R(x)=\Gamma_{\phi /l} \kappa (x)
 \end{equation}
where $\Gamma_{\phi /l}=\frac{V^\phi_m}{\Delta S_{\phi l}}\sigma_{\phi l}$ is the $\phi/l$ Gibbs Thomson coefficient.

As stated by Jackson and Hunt \cite{Jackson1966}, the mean undercooling at the interface can be computed on half of a eutectic period : 
\begin{equation}
\overline{\Delta T}=\frac{2}{\lambda}\int^{\lambda/2}_0 \Delta T(x) dx
\label{meandTgeneral}
\end{equation} 
As in Eq. (\ref{Tgeneral}), this mean eutectic undercooling can be separated as a mean solutal undercooling $\overline{\Delta T}_C$ and a mean curvature undercooling $\overline{\Delta T}_R$. Hillert \cite{Hillert1957}, and Jackson and Hunt \cite{Jackson1966} have shown that for interfaces that have constant mean curvature,   
\begin{equation}
\overline{\Delta T}_R=\frac{K_R}{\lambda} 
\label{curvature}
 \end{equation}
with 
\begin{equation}
K_R=2\left(\Gamma_{\alpha /l}\sin(\mid \theta_\alpha \mid)+\Gamma_{\beta /l}\sin(\mid \theta_\beta \mid) \right) 
\label{KR}
 \end{equation}
where angles $\theta_\alpha$ and $\theta_\beta $ are defined in the Figure \ref{fig:intro}.  

To define the mean solutal undercooling given in Equation (\ref{dTsolute})  requires an expression for the liquid concentration of the different elements at the interface. This necessitates a solution to the diffusion equation in the liquid phase for all independent $i$ elements: 
\begin{equation}
D_i \nabla^2 C^l_i + \vec{V} \cdot \vec{\nabla} C^l_i=0\;\; i=2,3 \ldots N
\end{equation}
where $D_i=\tilde{D}_{ii}$ is diagonal term of the interdiffusion coefficient matrix for element $i$. Here we neglect off diagonal terms since there is very little information on the magnitude or even the sign of these coefficients.  Solutions of this equation should satisfy the boundary conditions: 
\begin{eqnarray}
C_i=C^\infty_i\; \;&z\rightarrow \infty& \\
\frac{\partial C_i}{\partial x}=0\; \;&x=0,\frac{\lambda}{2} 
\end{eqnarray}

Therefore, the liquid concentration of any element $i$ can be expressed as:
\begin{equation}
C^l_i(x,z)=C^\infty_i+E^0_i\exp(-\frac{V}{D_i}z) +\displaystyle\sum_{n=1}^\infty E^n_i\exp(-\frac{2\pi n}{\lambda} z)\cos(\frac{2\pi n}{\lambda} x)
\label{Clgeneral}
\end{equation} 
for small Peclet numbers,  $Pe_i=\frac{V \lambda}{2 D_i}\ll 1$.

Assuming that all phases have the same molar volume, the conservation of matter at the interface gives for any element $i$:
\begin{equation}
D_i\frac{\partial C^l_i}{\partial z}\bigg|_{z=0}=V(C^s_i -C^l_i) \; \; \; i=1\dots N
\label{fluxinter}
\end{equation}
\subsection{Binary alloys: Jackson-Hunt-Hillert \label{sec:JHH}}
For binary alloys, Hillert \cite{Hillert1957} and Jackson and Hunt \cite{Jackson1966} used Eq. (\ref{fluxinter}) and the hypothesis of a constant concentration in the liquid phase 
at the interface to compute $E^n_i$ coefficients for $n>0$. 
In addition, for a microstructure similar to the one of Fig. \ref{fig:intro}, the solid/liquid interface could be reasonably supposed to be isothermal.
Using this hypotheses, Jackson and Hunt observed that the $E^0_i$ term of eq. (\ref{Clgeneral}) could be eliminated from the mean undercooling expression by using the relation:
\begin{equation}
\overline{\Delta T}^{iso}=\frac{m^\beta_2 \overline{\Delta T}^\alpha-m^\alpha_2 \overline{\Delta T}^\beta}{m^\beta_2-m^\alpha_2}
\label{trickJH}
\end{equation}
This approach masks the fact that the hypothesis of an isothermal interface gives a condition on the average liquid composition at the interface and so on the $E^0_i$ coefficients. Indeed, this growth condition requires in general a variation of the average liquid composition compared to the eutectic composition that equalizes the average undercooling of the two solid phases. As the interface is supposed to be at the thermodynamic equilibrium, this variation of composition in the liquid phase induces variations of compositions in solid phases and so an evolution of the solid fraction of phases compared to the one corresponding to phases composition at the eutectic temperature. One can note that inversely, a variation of phases solid fractions would induce variations of compositions in the solid phases and so a variation of the average liquid composition at the interface. 
 The relation used by Jackson and Hunt accounts these variations in the mean undercooling expression in an implicit way and avoids the computation of these variations.   Jackson and Hunt finally obtain an expression for the mean undercooling of an isothermal interface as a function of the growth velocity and the eutectic spacing:
\begin{equation}
\overline{\Delta T}^{iso}=K_1V\lambda+\frac{K_2}{\lambda} 
\label{dTisoJH}
\end{equation}
In addition, Jackson and Hunt observed that the eutectic spacing corresponding to the minimum undercooling ($\lambda_m$) satisfies the relation: 
\begin{equation}
\lambda_m^2 V=\frac{K_2}{K_1} 
\label{lambdam}
\end{equation}
This $\lambda_m$ is a scaling parameter of the microstructure developed at a given velocity. Although it has been shown that eutectics do not grow with a unique eutectic spacing at a given velocity, the microstructure developed is usually close to the one at $\lambda_m$. This is why eq. (\ref{lambdam}) is frequently used to characterize the microstructure developed by 2-phase eutectics. 

Unfortunately, the approach used by  Jackson and Hunt cannot be used for the N-component eutectic  growth problem. 
We  thus explicitly determine the general expression of the average concentration at the interface as a function of the volume fraction of the solid phases without any hypotheses on the undercooling and then compute the variation of the phase fractions corresponding to a shift of the average liquid concentration to make the interface isothermal. We finally determine the expression of the mean undercooling corresponding to the isothermal growth.  

\subsection{Approach}
We first determine the expression for the liquid concentration of all independent element $i$, the coefficients $E^0_i$ and $E^n_i$ (for $n>0$),  assuming that the Peclet number of any element $i$ ($Pe_i=\frac{V \lambda}{2 D_i}$) is small compared to 1. We then use the isothermal hypotheses to obtain an expression of the solid fraction variation with the undercooling, and finally express the mean undercooling of an isothermal interface as a function of the growth velocity and eutectic spacing. All of these steps imply a development of expressions at first order in Peclet numbers. For consistency, we therefore suppose that $\max ((Pe_2)^2,\dots,(Pe_N)^2)<\min (Pe_2,\dots,Pe_N)$ which implies that $\max (Pe_2,\dots,Pe_N)< \frac{\min (D_2,\dots,D_N)}{\max (D_2,\dots,D_N)}$.

\subsubsection{Liquid concentration field \label{sec:concentrations}}

\begin{figure}[h!]
\centering
\includegraphics[width=0.7\textwidth]{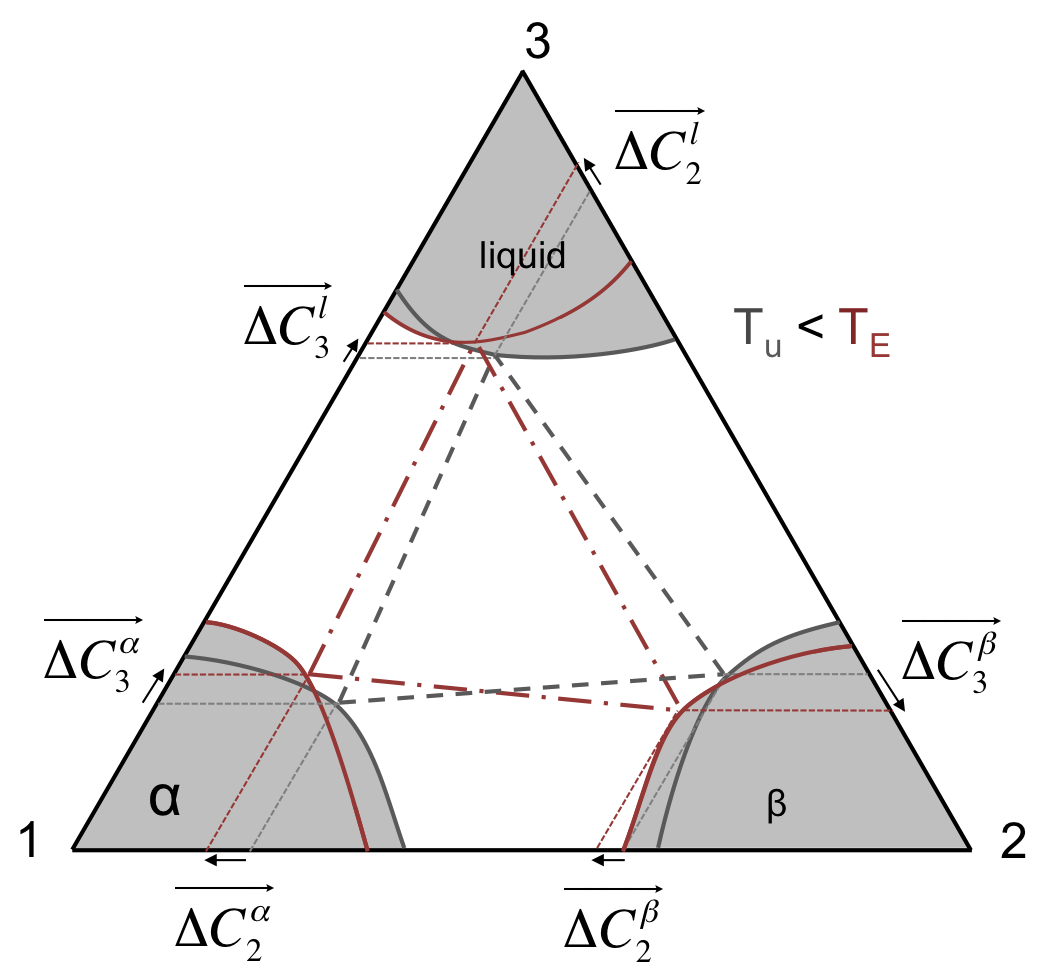}
\caption{\label{fig:tieline} Sketch of the evolution of equilibrium tie lines with temperature in a ternary two-phase eutectic. The arrows represent the concentration of elements 2 and 3 evolution in the different phases with the temperature evolution from $T_E$ to $T_u$}
\end{figure}

In this section we determine the coefficients $E^0_i$ and $E^n_i$, that are needed  in the general expression of the liquid concentration of element $i$ (Eq. \ref{Clgeneral}).
This entire analysis is performed at the solid-liquid interface, which corresponds to $z=0$. Therefore, the $z$-dependence of $C^l_i$ (see Eq. (\ref{Clgeneral})) does not appear in this section.
Introducing Eq. (\ref{Clgeneral}) in Eq. (\ref{fluxinter}) and expressing the function $C^l_i(x)-C^\phi_i(x)$ as a Fourier series we obtain  for $i=2\dots N$:
\begin{equation}
E^n_i=\frac{Pe_i}{\pi n}\frac{4}{\lambda}\int^{\lambda/2}_0(C^l_i(x)-C^s_i(x))\cos(\frac{2\pi n}{\lambda}x)dx  \; \; \; \mathrm{for}\,n>0
\label{Eni}
\end{equation}
So $E^n_i$ coefficients are at least first order  in the Peclet numbers. To proceed with the calculation of $E^n_i$ coefficients, a relationship between $C^l_i(x)$ and $C^s_i(x)$ is needed. For this, we use the relation obtained in Appendix \ref{Appendix A}
\begin{equation}
\Delta C^\phi_i(x)=\displaystyle \sum^N_{j=2} \Lambda^\phi_{ij}\Delta C^l_j(x) \label{coeffLambda}
\end{equation}
where $\Lambda^\phi_{ij}$ are certain solute distribution coefficients associated with the phase $\phi$ and assuming the effect of curvature on solid phase concentration can be neglected, see appendix \ref{Appendix A}. The $\Lambda^\phi_{ij}$ coefficients are functions of derivatives of chemical potentials that are function of concentration of elments and temperature. The full expression of $\Lambda^\phi_{ij}$ coefficients is presented in appendix \ref{Appendix A} for ternary alloys.
 This link between phase compositions is represented by a tie line in binary phase diagrams and by a tie triangle in isothermal cross section of  a ternary phase diagram. The tie triangle at equilibrium and that at an undercooling are shown in Figure \ref{fig:tieline}.  So as the liquid composition along the interface deviates from its equilibrium value, the tie triangle changes in shape, as given by the red dot-dash lines.  This change in shape is thus given by $(\overline{\Delta C_2}^\phi,..,\overline{\Delta C_N}^\phi)$ for a certain $(\overline{\Delta C_2}^l,..,\overline{\Delta C_N}^l)$

The approximation of $E^n_i$ to  first order in Peclet number implies that in Eq. (\ref{Eni}), $C^l_i(x)-C^s_i(x)$ needs only to be  approximated at zero order in Peclet number. Eqs (\ref{Clgeneral}) evaluated at $z=0$ and Eq. (\ref{coeffLambda}) give  for any position $x$ at the interface:
\begin{eqnarray}
C^l_i(x)\big|_{Pe^0_i}=C^\infty_i+E^0_i\big|_{Pe^0_i}\\
C^\phi_i(x)\big|_{Pe^0_i}-C^{\phi E}_i=\displaystyle \sum^N_{j=2} \Lambda^\phi_{ij} (C^l_j(x)\big|_{Pe^0_i}-C^{lE}_i)
\end{eqnarray}
where the index $Pe^0_i$ indicates that the expression is truncated at the zero order in Peclet numbers. We thus obtain that for both solid phases $\phi$:
\begin{equation}
(C^l_i(x)-C^\phi_i(x))\big|_{Pe^0_i}\simeq C^\infty_i+E^0_i\big|_{Pe^0_i}- \displaystyle\sum_{j=2}^N \Lambda^\phi_{ij}(C^\infty_j+E^0_j\big|_{Pe^0_j}-C^{lE}_j)-C^{\phi E}_i
\label{truncation0}
\end{equation}
 By introducing Eq. (\ref{truncation0}) in Eq. (\ref{Eni}) we get  for $n>0$:
\begin{equation}
E^n_i=\frac{V\lambda}{D_i}\frac{\sin(n\pi f_\alpha)}{(n\pi)^2}E_i
\label{Ein}
\end{equation}
where
\begin{equation}
E_i=\Delta C_i+\displaystyle\sum_{j=2}^N \Delta\Lambda_{ij}(C^\infty_j+E^0_j\big|_{Pe^0_j}-C^{lE}_j)
\label{Ei}
\end{equation}
with $\Delta C_i=C^{\beta E}_i-C^{\alpha E}_i$ and $\Delta\Lambda_{ij}=\Lambda^\beta_{ij}-\Lambda^\alpha_{ij}$. 
. 

It remains to determine the coefficient $E^0_i$.  For this, we use the fact that the the thermodynamic equilibrium concentration of element i in the liquid phase at the interface and at the eutectic temperature is $C^{lE}_i$. Therefore, if the volume fraction of solid phases does not change with undercooling below $T_E$, then the mean concentration of element $i$ the liquid phase at the interface is equal to $C^{lE}_i$, at zeroth order in the Peclet numbers.
 In this case, by integrating Eq. (\ref{Clgeneral}) over $\frac{\lambda}{2}$ we obtain:
\begin{eqnarray}
E^0_i(f^E_\alpha)\big|_{Pe^0_i}=C^{lE}_i-C^\infty_i\\
E_i(f^E_\alpha)=\Delta C_i
\end{eqnarray}
However, the volume fraction of phases can evolve with the undercooling and this evolution has to be introduced in the expression of elements concentration in the liquid phase. This is particularly important in the multicomponent alloy. 
For this, we use the conservation of matter between the solid phases and the liquid phase which implies that for each element $i$:
\begin{equation}
f_\alpha \overline{C^\alpha_i}+f_\beta \overline{C^\beta_i}=C^\infty_i
\label{massconserv}
\end{equation}
\begin{equation}
f^E_\alpha C^{\alpha E}_i+f^E_\beta C^{\beta E}_i=C^\infty_i
\label{massconservE}
\end{equation}
where $\overline{C^\alpha_i}$ (resp $\overline{C^\beta_i}$) is the average concentration of element $i$ in the solid phase $\alpha$ (resp $\beta$).
These two equalities imply that for each element $i=2\dots N$:
\begin{equation}
-\Delta f_\alpha \Delta C_i=f_\alpha(\overline{C^s_i}^\alpha-C^{\alpha E}_i)+f_\beta (\overline{C^s_i}^\beta-C^{\beta E}_i)
\label{Wi}
\end{equation}
where $\Delta f_\alpha=f^E_\alpha-f_\alpha$.
This system of equalities can be linked to variations of composition in the liquid phase using Eq. (\ref{coeffLambda}) averaged on the length of solid phases $\alpha$ and $\beta$.

The integration of Eq. (\ref{Clgeneral}) on each solid phase and using (\ref{Ein}) gives that for $i=2\dots N$:
\begin{eqnarray}
\overline{C^l_i}^\alpha=C^\infty_i+E^0_i+\frac{V\lambda}{D_i}\frac{1}{f_\alpha}E_i Q \label{Cilalpha}\\
\overline{C^l_i}^\beta=C^\infty_i+E^0_i-\frac{V\lambda}{D_i}\frac{1}{f_\beta}E_i Q \label{Cilbeta}
\end{eqnarray}
with $Q=\displaystyle\sum_{n=1}^\infty \frac{\sin^2(n\pi f_\alpha)}{(n\pi)^3}$. Introducing (\ref{coeffLambda}) and (\ref{Cilalpha}-\ref{Cilbeta}) into (\ref{Wi}) for each element $i$ leads to the system of equations for $E^0_i$:
\begin{equation}
\begin{bmatrix}
\bar{\Lambda}
\end{bmatrix}
\times
\begin{bmatrix} 
C^\infty_2+E^0_2-C^{lE}_2\\
\vdots\\
C^\infty_N+E^0_N-C^{lE}_N
\end{bmatrix}
=
\begin{bmatrix} 
-\Delta f_\alpha \Delta C_2\\
\vdots\\
-\Delta f_\alpha \Delta C_N
\end{bmatrix}
+
Q
\begin{bmatrix}
\Delta\Lambda
\end{bmatrix}
\times
\begin{bmatrix}
\frac{V\lambda}{D_2}E_2\\
\vdots\\
\frac{V\lambda}{D_N}E_N
\end{bmatrix}
\label{massLambda}
\end{equation}
where $\begin{bmatrix}\bar{\Lambda}\end{bmatrix}$ is the matrix of coefficients $\overline{\Lambda}_{ij}=f_\alpha \Lambda^\alpha_{ij}+f_\beta \Lambda^\beta_{ij}$ and $\begin{bmatrix}\Delta \Lambda\end{bmatrix}$ is the matrix of coefficients $\Delta \Lambda_{ij}$.
Solving Eq (\ref{massLambda}) for $E^0_i\big|_{Pe^0_i}$ along with Eq (\ref{Ei}) yields $E_i$ as a function of $\Delta f_\alpha, \Delta C_i$ and $\overline{\Lambda}_{ij}$ coefficients. 
We note that the $E_i$ coefficient is different from $\Delta C_i$ only if the phase fractions evolve compared to those at the eutectic temperature.
Using (\ref{massLambda}) enables us to obtain a full expression for the composition field for all $i$ independent concentrations in the liquid phase. We observe that the expression for the liquid phase concentration  depends on the volume fraction of solid phases as was discussed in part \ref{sec:JHH}.\\

Integrating Eqs (\ref{dTsolute}) and (\ref{dTcurvature}) on both solid phase interfaces, we obtain the mean undercooling of the $\alpha$ phase and of the $\beta$ phase have the expressions:
\begin{eqnarray}
\overline{\Delta T}^\alpha &=& \displaystyle\sum_{i=2}^N m^\alpha_i (C^{lE}_i-\overline{C^l_i}^\alpha) +\frac{2 \Gamma_{\alpha /l}\sin(\mid \theta_\alpha \mid)}{f_\alpha \lambda}\label{dTalpha}\\
\overline{\Delta T}^\beta &=& \displaystyle\sum_{i=2}^N m^\beta_i (C^{lE}_i-\overline{C^l_i}^\beta)+\frac{\Gamma_{\beta /l}\sin(\mid \theta_\beta \mid)}{f_\beta \lambda}\label{dTbeta}
\end{eqnarray}
where expressions of $\overline{C^l_i}^\alpha$ and $\overline{C^l_i}^\beta$ are given in Eqs (\ref{Cilalpha}) and (\ref{Cilbeta}). We thus observe that for a given growth velocity $V$ and eutectic spacing $\lambda$, $\overline{\Delta T}^\alpha $ and $\overline{\Delta T}^\beta $ yield different undercoling at each phase, given the phase fraction $f_\alpha$. This implies that in general, the values of $\overline{\Delta T}^\alpha $ and $\overline{\Delta T}^\beta $ evaluated at $f^E_\alpha$ can be very different as will be computed for binary alloys in section \ref{sec:Binary}. The growth of the eutectic at the velocity $V$ and eutectic spacing $\lambda$ with an isothermal interface therefore implies an evolution of solid fractions from $(f^E_\alpha,f^E_\beta)$. 
In the following, we compute the change in the phase fraction from that at equilibrium that is necessary to make the interface isothermal for a given growth velocity and eutectic spacing, which yields  the mean undercooling of the interface corresponding.

\subsubsection{Isothermal Interfaces \label{sec:dfiso}}
In this section, we determine the phase fraction variation induced by requiring an isothermal interface for a given growth velocity and eutectic spacing. For this, the interface is  isothermal if the mean undercoolings of the $\alpha$ phase and the $\beta$ phase are equal: 
\begin{equation}
\overline{\Delta T}^\alpha(f_\alpha)=\overline{\Delta T}^\beta(f_\alpha)
\end{equation}
A Taylor expansion of this equality to first order in the variation of $f_\alpha$ gives:
\begin{equation}
\Delta f^\mathrm{iso}_\alpha\left(\frac{\partial \overline{\Delta T}^\beta}{\partial f_\alpha}\bigg|_{f^E_\alpha}- \frac{\partial \overline{\Delta T}^\alpha}{\partial f_\alpha}\bigg|_{f^E_\alpha}\right) = \overline{\Delta T}^\beta(f^E_\alpha)-\overline{\Delta T}^\alpha(f^E_\alpha)
\label{eq:dfiso1}
\end{equation}
with $\Delta f^\mathrm{iso}_\alpha=f^E_\alpha-f^\mathrm{iso}_\alpha$ where $f^\mathrm{iso}_\alpha$ is the solid fraction corresponding to the undercooling of an isothermal interface.

Using the expressions for the undercoolings of each phase (\ref{dTalpha}) and (\ref{dTbeta}), and the expressions for  the mean liquid concentrations at each phase interface (\ref{Cilalpha}) and (\ref{Cilbeta}) we obtain:
\begin{equation}
\overline{\Delta T}^\beta(f^E_\alpha)-\overline{\Delta T}^\alpha(f^E_\alpha)=V\lambda \alpha_C+\frac{\alpha_R}{\lambda}
\label{dfnumerator}
\end{equation}
where
\begin{equation}
\alpha_R=2\left[\frac{\Gamma_{\beta /l}\sin(\mid \theta_\beta \mid)}{f^E_\beta}- \frac{\Gamma_{\alpha /l}\sin(\mid \theta_\alpha \mid)}{f^E_\alpha}\right] 
\label{alphaR}
\end{equation}
and $\alpha_C$ will be given for binary (section \ref{sec:Binary}) and ternary (section \ref{sec:Ternary}) eutectics.  
 We rename for simplicity 
\begin{equation}
-\Delta'=\frac{\partial \overline{\Delta T}^\beta}{\partial f_\alpha}\bigg|_{f^E_\alpha}- \frac{\partial \overline{\Delta T}^\alpha}{\partial f_\alpha}\bigg|_{f^E_\alpha}
\label{Deltadef}
\end{equation}
We assume for the following that $\Delta'\approx \Delta'_0$ where $\Delta'_0$ is independent on velocity and the eutectic spacing.
The validity of this hypotheses is discussed in appendix \ref{Appendix B}. This assumes that the influence of a variation of solid fractions on $\overline{\Delta T}^\alpha$ and $\overline{\Delta T}^\beta$ results only from a change of the average composition of the liquid phase at the interface.

Using (\ref{dfnumerator}) and (\ref{Deltadef}) in (\ref{eq:dfiso1}) we thus get the expression of the variation of $f_\alpha$ necessary to yield an isothermal interface:
\begin{equation}
\Delta f^\mathrm{iso}_\alpha=-\left(V\lambda\frac{\alpha_C}{\Delta'_0}+\frac{1}{\lambda}\frac{\alpha_R}{\Delta'_0} \right)
\label{dfisoexp}
\end{equation}
The general expression of $\Delta'_0$ for a given phase diagram   is given in appendix \ref{Appendix B}. For binary alloys, this  expression gives  $\Delta'_0={\Delta m_2 \Delta C_2}/{\overline{\Lambda_{22}}^E}$ where $\Delta m_2=m^\beta_2-m^\alpha_2$ which is always positive. This means that if $\overline{\Delta T}^\beta(f^E_\alpha)>\overline{\Delta T}^\alpha(f^E_\alpha)$ the fraction of $\alpha$ phase has to be increased to make the interface isothermal and if $\overline{\Delta T}^\beta(f^E_\alpha)<\overline{\Delta T}^\alpha(f^E_\alpha)$ the fraction of $\beta$ phase has to be increased to make the interface isothermal, which makes sense intuitively, as already discussed by Magnin and Trivedi \cite{Magnin1991}.

\subsubsection{Undercooling of isothermal interface \label{sec:meanUndercooling}}

We now determine the expression for the mean undercooling of the isothermal interface.
The mean undercooling defined in Eq. (\ref{meandTgeneral}) can be computed using Eqs. (\ref{curvature}) and (\ref{dTsolute}),  as a function of the volume fraction of phases using the liquid concentrations  obtained in section \ref{sec:concentrations}. For small changes of the volume fractions of the solid phases compared to their equilibrium values, $(f_\alpha^E,f_\beta^E)$, the mean undercooling can be approximated by a Taylor expansion to first order in the change of $f_\alpha$ from $ f_\alpha^E$. Moreover, we have seen in section \ref{sec:dfiso} that for a given growth velocity and eutectic spacing, the system enforces an isothermal condition by changing the average concentration at the interface which corresponds to a variation of phases fractions $\Delta f^\mathrm{iso}_\alpha$. 

 We can thus express the mean undercooling of an isothermal interface as:

\begin{equation}
\overline{\Delta T}^\mathrm{iso}(f_\alpha)=\overline{\Delta T}_C(f^E_\alpha)-\Delta f^\mathrm{iso}_\alpha \frac{\partial \overline{\Delta T}_C}{\partial f_\alpha}\bigg|_{f^E_\alpha}+\overline{\Delta T}_R
\label{dTmeanIso}
\end{equation}
In order to use the expression of $\overline{\Delta T}^\mathrm{iso}$  given in (\ref{dTmeanIso}), we need an expression for $\overline{\Delta T}_C(f^E_\alpha)$ and $\frac{\partial \overline{\Delta T}_C}{\partial f_\alpha}\bigg|_{f^E_\alpha}$.  
From section \ref{sec:concentrations} we obtain : 
\begin{equation}
\overline{\Delta T}_C(f^E_\alpha)=\displaystyle\sum_{i=2}^N (C^{lE}_i-C^\infty_i-E^0_i\big|_{f^E_\alpha})\overline{m_i}^E+\displaystyle\sum_{i=2}^N \frac{V \lambda}{D_i}(E_i Q)\big|_{f^E_\alpha}\Delta m_i 
\end{equation}
The system of equations for $E^0_i$ (\ref{massLambda}) shows that for $i=2\dots N$, $C^{lE}_i-C^\infty_i-E^0_i\big|_{f^E_\alpha}$ is proportional to $Pe_i$, so we can write:
\begin{equation}
\overline{\Delta T}_C(f^E_\alpha)=V \lambda K_C
\label{dTfe}
\end{equation}
The expression of $K_C$ will be given for binary and ternary alloys in sections \ref{sec:Binary} and \ref{sec:Ternary} respectively.
For the term involving $\frac{\partial \overline{\Delta T}_C}{\partial f_\alpha}\bigg|_{f^E_\alpha}$ in Eq. (\ref{dTmeanIso}), it is unclear if this quantity has to be evaluated at first order in Peclet numbers or at zero order, as the order of $\Delta f^\mathrm{iso}_\alpha $  in Peclet has not been determined.
For simplicity, we approximate $\frac{\partial \overline{\Delta T}_C}{\partial f_\alpha}\bigg|_{f^E_\alpha}$ at zero order in Peclet number. This hypothesis is discussed in appendix \ref{Appendix B} by analyzing the range of order of $\Delta f^\mathrm{iso}_\alpha$. Introducing Eq. (\ref{dTfe}) and (\ref{dfisoexp}) in (\ref{dTmeanIso}) we thus obtain the undercooling of the isothermal interface:
\begin{equation}
\overline{\Delta T}^\mathrm{iso}=V\lambda K_1+\frac{K_2}{\lambda}
\label{dTiso1}
\end{equation}
where $K_1$ and $K_2$ coefficients are:
\begin{eqnarray}
K_1=K_C+l_N \alpha_C \label{K1exp}\\
K_2=K_R+l_N \alpha_R  \label{K2exp}
\end{eqnarray}
and 
\begin{equation}
l_N=\frac{1}{\Delta'_0}\frac{\partial \overline{\Delta T}_C}{\partial f_\alpha}\bigg|_{f^E_\alpha,Pe_i^0}
\label{lN}
\end{equation}
For a given growth velocity, we thus have now established the link between the mean temperature at the isothermal interface and the eutectic spacing for any 2-phase eutectic with N elements. 

From eq (\ref{dTiso1}) we obtain that the eutectic spacing corresponding to the minimum undercooling verifies the relation:
\begin{equation}
\lambda_m^2 V=\frac{K_2}{K_1}
\end{equation}
These expressions show that the growth law (\ref{dTisoJH}) determined by Jackson and Hunt \cite{Jackson1966} for binary alloys can be generalized to  two-phase eutectics with N-elements. However, the analytical expressions for $K_C$, $l_N$ and $\alpha_C$  can quite complicated with $N$ large. The thermodynamic  parameters needed  to evaluate these coefficients can be found using CALPHAD descriptions of the free energies.  These coefficients are therefore only given here for binary alloys (in section \ref{sec:Binary}) and for ternary alloys (in section \ref{sec:Ternary}). In a similar way as the Jackson Hunt theory, this general model takes into account interfacial energies of the two solid phases  through the expression of the undercooling of the isothermal interface through Gibbs-Thomson coefficients ($\Gamma_{\alpha l}$, $\Gamma_{\beta l}$) and  trijunction angles  ($\theta_\alpha$, $\theta_\beta$).
The diffusion properties of the alloy are also introduced in the theory through the interdiffusion coefficients of each independent element $\left\lbrace D_2,\dots,D_N \right\rbrace $. As in the Jackson Hunt model, this theory includes thermodynamic properties of the alloy which correspond to the equilibrium concentration of elements in solid phases ($C^{\alpha E}_i$, $C^{\beta E}_i$), the liquidus slopes corresponding to each phase ($m^\alpha_i$ and $m^\beta_i$) and certain solute distribution coefficients ($\left[ \Lambda^\alpha \right] $ and $\left[ \Lambda^\beta \right] $) defined in Appendix \ref{Appendix A}. The expressions of liquidus slopes and solute distribution coefficients as functions of derivatives of chemical potentials for a given  temperature and  concentration are given in Appendix \ref{Appendix A}. These derivatives can be computed from the expression of the Gibbs free energy of phases using CALPHAD descriptions of the free energies.

Catalina et al. \cite{Catalina2015} have recently proposed a model for the growth of two-phase eutectics with N elements in the limit that  the composition of the $\beta$ phase is a constant $[\Lambda^\beta]=0$. Moreover, this model only takes into account diagonal terms of the $[\Lambda^\alpha]$ matrix. Our  model is thus a generalization of this approach.  
To illustrate the predictions of the model, we examine the coefficients $K_1$ and $K_2$ for binary and ternary alloys.  

\section{Binary alloys \label{sec:Binary}}

In this section, we illustrate how the general theory can be used to describe the well-known results in a binary alloy. From the development of the solute concentration expression at the interface given in section \ref{sec:concentrations} we obtain the concentration at the interface:
\begin{equation}
C^l_2(x)=C^\infty_2+E^0_2+\frac{V \lambda}{D_2}E_2 \displaystyle\sum_{n=1}^\infty \frac{\sin (n \pi f_\alpha)}{(n \pi)^2} \cos (\frac{2 \pi}{\lambda}x)
\end{equation}
with
\begin{eqnarray}
E^0_2&=&C^{lE}_2-C^\infty_2-\frac{\Delta C_2}{\overline{\Lambda}_{22}}\Delta f_\alpha+Q\frac{\Delta \Lambda_{22}}{\overline{\Lambda}_{22}}\frac{V \lambda}{D_2}E_2 \label{E02}\\
E_2&=&\Delta C_2\left(1- \frac{\Delta \Lambda_{22}}{\overline{\Lambda}_{22}}\Delta f_\alpha \right) \label{E2}
\end{eqnarray}
Donaghey and Tiller \cite{Donaghey1968} give a detailed development at first order in Peclet number of the concentration in the liquid phase for binary alloys. Our expressions for the  parameters $E^0_2$ and $E_2$ defined in equations (\ref{E02}) and (\ref{E2}) are identical the one obtained by Donaghey and Tiller \cite{Donaghey1968}.

This expression for the solute concentration at the interface can be introduced in Eq. (\ref{dTsolute}) to determine the solutal undercooling at any position $x$ of the interface. 
The integration of $\Delta T_C (x)$ on half of the eutectic spacing gives coefficients $K_C$ and the $\frac{\partial \overline{\Delta T}_C}{\partial f_\alpha}\bigg|_{f^E_\alpha,Pe_2^0}$ term in $l_2$ (see Eq. (\ref{lN})) and its integration on each solid phase interface gives the coefficient $\alpha_C$ (see eq. (\ref{dfnumerator})) and the $\Delta'_0$ term in $l_2$ (see Eq. (\ref{lN})) and obtains: 
\begin{eqnarray}
l_2&=&-\frac{\overline{m}_2^E}{\Delta m_2} \label{l2}\\
K_C&=&Q^E \frac{\Delta C_2}{D_2}\left[\Delta m_2-\frac{\Delta \Lambda_{22}}{\overline{\Lambda_{22}}^E}\overline{m}_2^E \right] \label{KC2}\\
\alpha_C&=&Q^E \frac{\Delta C_2}{D_2}\left[-\frac{\Delta \Lambda_{22}}{\overline{\Lambda_{22}}^E}\Delta m_2+\left(\frac{m^\beta_2}{f^E_\beta}+\frac{m^\alpha_2}{f^E_\alpha} \right)  \right] \label{alphaC2}
\end{eqnarray}
where $\overline{m}_2=f_\alpha m^\alpha_2+f_\beta m^\beta_2$. The coefficients $m^\phi_i$ are signed and so, for binary alloys, the $m^\alpha_2$ coefficient is negative and the solute distribution coefficients $\Lambda^\alpha_{22}$ and $\Lambda^\beta_{22}$ are usually noted $k^\alpha$ and $k^\beta$ for binary alloys. In arriving at these results, we employ the result that follows from  Eqs. (\ref{eqMatrix}) and Eq. (\ref{chemcoeff}) in appendix \ref{Appendix A}  that relates the second derivates of the free energy to the slope of the liquidus:
\begin{equation}
m^\phi_2=-\frac{ \left( C^{lE}_2-C^{\phi E}_2 \right) \frac{\partial^2 G^l_m}{\partial (C^l_2)^2} }{\Delta S_{\phi l}}
\label{mlbinary}
\end{equation}
where $G^l_m$ is the molar Gibbs free energy of the liquid phase. This expression for the slope of the phase $\phi$ liquidus curve is the well-known Gibbs-Konovalov  relation \cite{Goodman1981}. 

By introducing Eqs. (\ref{l2}), (\ref{KC2}) and (\ref{alphaC2}) in expressions of coefficients $K_1$ and $K_2$ (eqs. (\ref{K1exp}) and (\ref{K2exp})) we obtain that :
\begin{eqnarray}
K_1&=&\left(\frac{-m^\alpha_2 m^\beta_2}{\Delta m_2 f^E_\alpha f^E_\beta} \right) Q^E \frac{\Delta C_2}{D_2}\\
K_2&=&\frac{-2 m^\alpha_2 \Gamma_{\beta /l}\sin(\mid \theta_\beta \mid)}{f^E_\beta \Delta m_2}+\frac{2 m^\beta_2 \Gamma_{\alpha /l}\sin(\mid \theta_\alpha \mid)}{f^E_\alpha \Delta m_2}
\end{eqnarray}
The $K_1$ and $K_2$ coefficients are identical to those obtained by Jackson and Hunt \cite{Jackson1966}. The coefficients  that set the $\lambda_m^2 V$ relationship should indeed  be the same as those of Jackson and Hunt, since   the same hypotheses and  approximations are used in our approach and were also used by Jackson and Hunt. However, our treatment yields the expression for the $E^0_2$ coefficient, and thus we can determine the effects of the asymmetry of the phase diagram on the volume fraction of the phases. 

Magnin and Trivedi \cite{Magnin1991}  published a eutectic growth model similar to ours for binary alloys. In their study, they determined the expression of the liquid concentration at the interface by using the conservation of matter at the interface (\ref{fluxinter}) and  
taking into account density differences between phases. They obtain the same expression for the mean undercooling as Jackson and Hunt, and $K_1$ and $K_2$ coefficients are identical with ours (Eqs (\ref{K1exp}) and (\ref{K2exp})), in the limit where the density of the phases are identical. However, our $K_C$ and $\alpha_C$ coefficients are different, as Magnin and Trivedi did not take into account the terms at first order in Peclet number in their $E^0_2$ parameter (the last term in the expression of $E^0_2$ in Eq. (\ref{E02})).
We showed above that if the interface is isothermal and undercooled then the fraction of the phases can change from their equilibrium values.  To illustrate this for a binary alloy,  we examine the difference of undercooling between the two solid phases if the phase fractions do not change with the growth conditions and thus the interface is nonisothermal.  
From eq. (\ref{dfnumerator}) we observe that $|\overline{\Delta T}^\beta(f^E_\alpha)-\overline{\Delta T}^\alpha(f^E_\alpha)|$ is a function of $\lambda$ at a given velocity. If $\alpha_R$ and $\alpha_C$ have the same sign, then $ |\overline{\Delta T}^\beta(f^E_\alpha)-\overline{\Delta T}^\alpha(f^E_\alpha)|$ has a minimum with the expression: 
\begin{equation}
\left( \overline{\Delta T}^\beta(f^E_\alpha)-\overline{\Delta T}^\alpha(f^E_\alpha)\right)\big|_{min}=2\sqrt{\alpha_C \alpha_R V}
\end{equation}
From Eq. (\ref{alphaR}) and (\ref{alphaC2}) we observe that coefficients $\alpha_R$ and $\alpha_C$ are large if the two solid phases have asymetrical properties and a low diffusion coefficient. For example, taking the following properties: $\Delta C_2=90\%$ and $D_2=5\times 10^{-10}\,\mathrm{m^2/s}$, $f_\alpha=0.2$, $ \Gamma_{\alpha /l}\sin(\mid \theta_\alpha \mid)=1\times 10^{-7}\,\mathrm{K.m}$, $ \Gamma_{\beta /l}\sin(\mid \theta_\beta \mid)=1\times 10^{-8}\,\mathrm{K.m}$, $m^\alpha_2=-50\,\mathrm{K.at\%}$, $m^\beta_2=5\,\mathrm{K.at\%}$, $\Lambda^\alpha_{22}=0.1$, $\Lambda^\beta_{22}=0.2$ we get that for $V=100\,\mathrm{\mu m /s}$ $\left( \overline{\Delta T}^\beta(f^E_\alpha)-\overline{\Delta T}^\alpha(f^E_\alpha)\right)\big|_{min}=17.7\,K$. For a standard thermal gradient $G=9\,\mathrm{K/mm}$, this difference of undercooling would induce a difference of position of $2\,\mathrm{mm}$ between the $\alpha /l$ and the $\beta /l$ interfaces which would be observable if the eutectic was not growing with an isothermal interface.
From these properties and using the expression of $\Delta'_0$ given in appendix \ref{Appendix B}, we compute that the change in  the $\alpha$ phase fraction needed to insure an isothermal interface is $\Delta f^\mathrm{iso}_\alpha=-1.3\times 10^{-3}$. Such a small variation of solid fraction would certainly be difficult to observe in experiments. However, other choices of materials parameters may yield larger changes.
If $\alpha_C$ and $\alpha_R$ do not have the same sign then, at a given velocity, the function $|\overline{\Delta T}^\beta(f^E_\alpha)-\overline{\Delta T}^\alpha(f^E_\alpha)|$  has a zero value for a ceryain $\lambda_0$. For this $\lambda_0$ the interface is isothermal at $f_\alpha=f^E_\alpha$. However, for eutectic spacings far from this $\lambda_0$ the difference of undercooling in front of the two solid phases can be very different for $f_\alpha=f^E_\alpha$. 

\section{Ternary alloys \label{sec:Ternary}}

We now apply the general method to ternary two-phase eutectics. The coefficients used in the theory are given and compared to those in binary alloys. The ternary model is compared to previous models available in the literature. Finally, we evaluate the evolution of the $\lambda_m^2 V$ law for a binary alloy with the addition of a small amount of element 3, when the system stays in a two-phase eutectic microstructure.

\subsection{General model in ternary alloys}

Using the same process as described in section \ref{sec:Binary}, we obtain that for ternary alloys, the coefficients $l_3$, $K_C$ and $\alpha_C$ defined in section \ref{sec:Theory} are:
\begin{equation}
l_3=-\frac{\Delta C_2\left(\overline{m_2}^E \overline{\Lambda_{33}}^E-\overline{m_3}^E\overline{\Lambda_{32}}^E\right)+\Delta C_3\left(-\overline{m_2}^E\overline{\Lambda_{23}}^E+\overline{m_3}^E\overline{\Lambda_{22}}^E\right)}{\Delta C_2\left(\Delta m_2 \overline{\Lambda_{33}}^E- \Delta m_3\overline{\Lambda_{32}}^E\right) + \Delta C_3\left(-\Delta m_2\overline{\Lambda_{23}}^E+\Delta m_3\overline{\Lambda_{22}}^E\right)}
\label{l3}
\end{equation}

\begin{eqnarray}
&K_C&=Q^E \displaystyle \sum^3_{i=2} \frac{\Delta C_i}{D_i}\left[\Delta m_i-\overline{m_2}^E\left(\frac{\Delta \Lambda}{\overline{\Lambda}} \right)^E_{2i} -\overline{m_3}^E\left(\frac{\Delta \Lambda}{\overline{\Lambda}} \right)^E_{3i} \right]  \label{KC3}\\
&\alpha_C &= Q^E \displaystyle \sum^3_{i=2} \frac{\Delta C_i}{D_i}\left[-\Delta m_2 \left(\frac{\Delta \Lambda}{\overline{\Lambda}} \right)^E_{2i}-\Delta m_3 \left(\frac{\Delta \Lambda}{\overline{\Lambda}} \right)^E_{3i}+\left(\frac{m^\beta_i}{f^E_\beta}+\frac{m^\alpha_i}{f^E_\alpha} \right) \right] \label{alphaC3}
\end{eqnarray}

where
\begin{eqnarray}
\left(\frac{\Delta \Lambda}{\overline{\Lambda}} \right)^E_{2i}=\frac{\overline{\Lambda}^E_{33} \Delta \Lambda_{2i}- \overline{\Lambda}^E_{23} \Delta \Lambda_{3i} }{\overline{\Lambda}^E_{22}\overline{\Lambda}^E_{33}-\overline{\Lambda}^E_{32}\overline{\Lambda}^E_{23}}\\
\left(\frac{\Delta \Lambda}{\overline{\Lambda}} \right)^E_{3i}=\frac{\overline{\Lambda}^E_{22}\Delta \Lambda_{3i}-\overline{\Lambda}^E_{32}\Delta \Lambda_{2i}}{\overline{\Lambda}^E_{22}\overline{\Lambda}^E_{33}-\overline{\Lambda}^E_{32}\overline{\Lambda}^E_{23}}
\end{eqnarray}

We observe that coefficients $l_3$, $K_C$ and $\alpha_C$ obtained for ternary alloys have the same form as coefficients obtained for binary alloys presented in Eq. (\ref{l2}-\ref{alphaC2}). However,  whereas $l_2$ only depends on liquidus slopes $m^\alpha_2$ and $m^\beta_2$ and on the phase fractions, coefficient $l_3$ also depends on differences of concentration in solid phases $\Delta C_2$ and $\Delta C_3$ and on the $\overline{\Lambda}_{ij}$ coefficients. 

McCartney and Hunt  \cite{McCartney1980} assume that the ratio $(E^0_3+C^\infty_3-C^E_3)/(E^0_2+C^\infty_2-C^E_2)$ is independent of the conditions for  eutectic growth. 
In addition, they assume that the non-diagonal terms in the $\left[ \bar{\Lambda}\right]$ matrix are negligible compared to diagonal terms and that $\overline{\Lambda}^E_{22} \simeq \overline{\Lambda}^E_{33}$, which limits this model to systems with specific thermodynamic properties. 
To illustrate this statement, we have computed the $\left[ \bar{\Lambda}\right]$ matrix in the ternary eutectic Al-Cu-Ag at the composition: $14.8\mathrm{at}\%Cu-5\mathrm{at}\%Ag$. For this, we have used the expression of the Gibbs free energies for the liquid phase, the $\alpha$ phase and the $\theta-Al_2Cu$ phase given in Ref. \cite{Witusiewicz2004,Witusiewicz2005} and computed the equilibrium composition of each phase at the eutectic temperature using ThermoCalc. We have obtain $\left[ \bar{\Lambda}\right]=\begin{bmatrix} 
0.46 & 0.20\\
0.48 & 0.94
\end{bmatrix}$ . Therefore, in this case, non-diagonal terms are of similar order to diagonal terms and that $\overline{\Lambda}^E_{22} \neq \overline{\Lambda}^E_{33}$. 
 Finally McCartney et al. have used the assumption that $D_2=D_3$ to obtain their final expression of the interface undercooling. Recently, DeWilde et al. \cite{DeWilde2005} proposed a new model for the directional growth of ternary two-phase eutectics. In this model, the mean solutal undercooling of each solid phase is expressed as a sum of absolute values of undercoolings corresponding to each element. In addition, this model neglects the dependence of $E^0_i$ on the change of the phase fractions with the undercooling. This approximation  eliminates the $\Delta'_0$ term in the expression of $\Delta'$ given Eq. (\ref{Deltapexp}) but  keeps the $V \lambda \xi_C+\frac{\xi_R}{\lambda}$ term (see Appendix \ref{Appendix B}). 
 We note that none of these assumptions made by McCartney et al. or DeWilde et al. are used in our theory. 
 
Some binary eutectics stay in a two-phase microstructure with the addition of a ternary element. In this case, if all parameters involved in coefficients $K_1$ and $K_2$ are known for the ternary alloy, one could predict the evolution of the microstructure with the addition of the element $3$ at a given velocity by comparing ${K_2}/{K_1}$ ratios of the binary and the ternary alloys using Eq. (\ref{lambdam}).
In the general case, this comparison is difficult due to the large number of parameters involved in these ratios. 
In particular, the equilibrium of the ternary eutectic might take place at a different temperature than the binary system which would affect all parameters involved in the growth law that depend on temperature such as the interfacial energies, and diffusion coefficients. From Eqs. (\ref{KC3}) and (\ref{alphaC3}), we see that if element 3 is a slow diffuser compared to element 2, the coefficients $K_C$ and $\alpha_C$ are  particularly sensitive to the thermodynamic parameters associated with element 3 and so the eutectic microstructure  might change drastically compared to the binary alloy.
We also note that if the solubility of element 3 is the same in the 2 phases and if the cross coefficients of the $[\Lambda^\phi]$ matrices are negligible, the coefficients $K_1$ and $K_2$ of the ternary alloy have the same form as the one of the binary alloy. 
Therefore, if the addition of component 3 does not affect element 2 thermodynamic ($ \Lambda^\phi_{22}$ and $m^\phi_2$ with $\phi=\alpha,\beta$) and diffusion ($D_2$) coefficients,
 then element 3 have no effect on the alloy eutectic spacing.

\subsection{Limit at low addition of a third element \label{lowAdd}}

\begin{figure}[h!]
\centering
\includegraphics[scale=0.4]{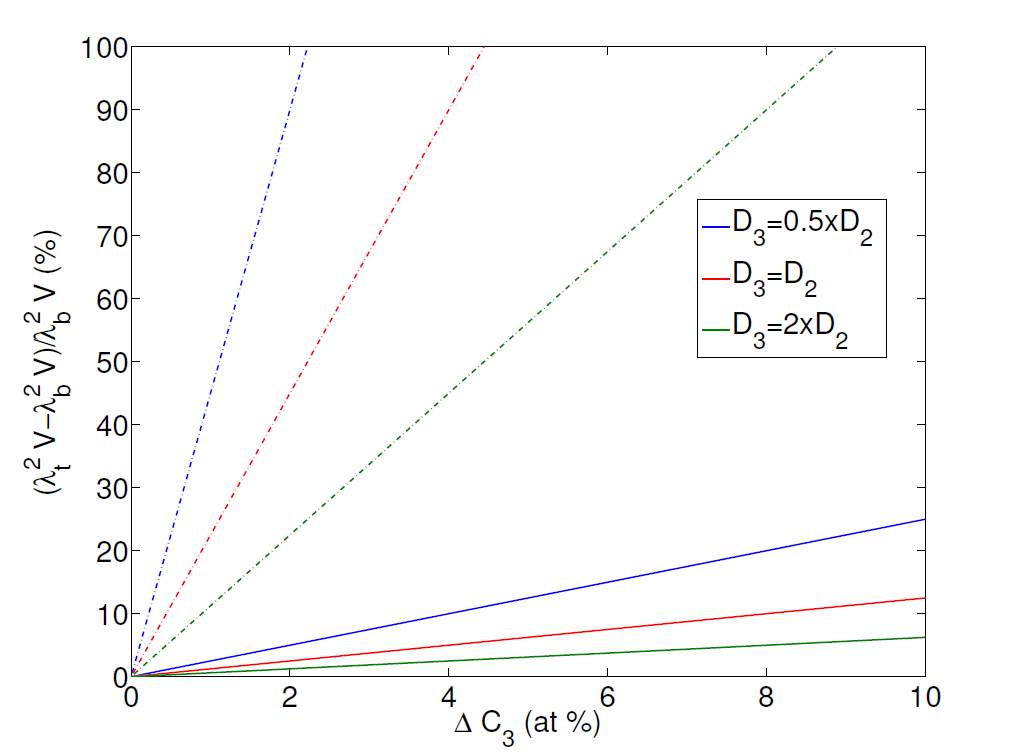}
\caption{\label{fig:symetrical} Variation of the $\lambda_m^2 V$ law with the addition of element 3 compared to the $\lambda_m^2 V$ law of the symmetrical binary alloy (in \%) according to $\Delta C_3$. Dashed lines correspond to: $m^\beta_3=-9\,K/(at\%)$ and $m^\alpha_3=-10\,K/(at\%) $, so $\mid \Delta m_3 \mid=1\,K/(at\%)$ and solid lines correspond to: $m^\beta_3=10\,K/(at\%)$ and $m^\alpha_3=-10\,K/(at\%) $, so $\mid \Delta m_3 \mid=20\,K/(at\%)$}
\end{figure}

To illustrate the effects of component 3 on the growth law $\lambda_m^2 V=\frac{K_2}{K_1}$ of a binary alloy, we consider the limit of a ternary allow with a dilute amount of component 3.  We assume that the capillary lengths, and the phases fractions do not change significantly with the addition of component 3. Coefficients $K_R$ and $\alpha_R$ are therefore  identical to those of a binary alloy. Moreover, the variation of the thermodynamic properties of element 2 ($ \Lambda^\phi_{22}$ and $m^\phi_2$ with $\phi=\alpha,\beta$) with the addition of element 3 is neglected.
In addition, for a low addition of element 3, we should have $|\Delta C_3|\ll |\Delta C_2|$. To simplify the problem, we also assume that the  $\Lambda^\phi_{23}$ and $\Lambda^\phi_{32}$ coefficients for the $\alpha$ and  $\beta$ phases are negligible. In this case, coefficients $K_1$ and $K_2$ of the ternary alloy growth law can be expressed as:
\begin{eqnarray}
K^t_1=K^b_1+\Delta C_3 q_1\\
K^t_2=K^b_2+\Delta C_3 q_2
\end{eqnarray}
where the 'b' exponent refers to the binary alloy and the 't' exponent refers to the ternary alloy and
\begin{eqnarray}
&q_1 &=Q^E \left\lbrace \frac{1}{D_3}\left[-\frac{m^\alpha_3 m^\beta_3}{\Delta m_3 f_\alpha f_\beta} \right]  + \frac{1}{D_2}\left(\frac{m^\alpha_2m^\beta_3-m^\alpha_3m^\beta_2}{(\Delta m_2)^2 } \right) \left[\frac{f_\alpha \Lambda^\alpha_{22}m^\beta_2+f_\beta\Lambda^\beta_{22}m^\alpha_2}{ f_\alpha f_\beta \overline{\Lambda_{33}}^E} \right]  \right\rbrace \label{eps1}\\
&q_2 &=\frac{1}{\Delta C_2}\frac{\overline{\Lambda_{22}}^E}{\overline{\Lambda_{33}}^E}\left( \frac{m^\alpha_2m^\beta_3-m^\alpha_3m^\beta_2}{(\Delta m_2)^2}\right) \alpha_R \label{eps2}
\end{eqnarray}
and so 
\begin{equation}
\lambda_{mt}^2 V=\frac{K^b_2}{K^b_1}+\frac{\Delta C_3}{K^b_1}\left( q_2-\frac{K^b_2}{K^b_1} q_1 \right) 
\label{l2Vlowadd}
\end{equation}
We observe in these equations that $q_1$ and $q_2$ depend on component 3 through the parameters $D_3$, $ \overline{\Lambda_{33}}^E$, $m^\alpha_3 $ and $m^\beta_3 $.
In these expressions, the thermodynamic coefficients $ \overline{\Lambda_{33}}^E$, $m^\alpha_3 $ and $m^\beta_3 $ can be determined from the Gibbs free energies of the phases  as shown in appendix \ref{Appendix A}.
If $m^\alpha_3$ and/or $m^\beta_3$ is small compared to other slopes, then the $q_1$ coefficient will be insensitive to $D_3$. So changes on liquidus curves of both solid phases with the addition of element 3 are necessary conditions for $D_3$ to have an effect on the eutectic spacing. \\
In the general case, there are many factors that lead to a change in $\lambda_m^2 V$ with the addition of a third element as shown in Eqs (\ref{eps1}-\ref{eps2}). However, if $f_\alpha=0.5$ and the phase diagram of the binary alloy is symmetrical ($m^\beta_2=-m^\alpha_2$ and $ \Lambda^\beta_{22}=\Lambda^\alpha_{22}$), then the term depending on $D_2$ in eq. (\ref{eps1}) disappears. In addition, if the binary alloy has equal solid/liquid surface energies for the $\alpha$ and the $\beta$ phase, then $q_2$ can be neglected and eq. (\ref{l2Vlowadd}) becomes:
\begin{equation}
\lambda_{mt}^2 V\big|_{symmetrical}=\frac{K^b_2}{K^b_1}\left(1+\frac{\Delta C_3}{\Delta C_2}\frac{D_2}{D_3}\frac{2 m^\beta_3}{m^\beta_2}\frac{m^\alpha_3}{\Delta m_3} \right) 
\label{l2Vsym}
\end{equation}
For this particular case, the evolution of $\lambda_m^2 V$ with the addition of element 3 can be analyzed according to element 3 parameters. The concentration of element 3 is given in terms of $\Delta C_3$ which we take to be positive. In fig. \ref{fig:symetrical}, we present the variation of the $\lambda_m^2 V$ according to $\Delta C_3$ for two different sets of $m^\alpha_3$ and $m^\beta_3$ coefficients and for different diffusion coefficients $D_3$. In this figure, parameters used for element 2 are: $D_2=1\times 10^{-9}\,\mathrm{m^2/s}$, $\Delta C_2=80\,(at\%)$ and $m^\beta_2=10\,K/(at\%)$.

We note from Eq. (\ref{l2Vsym}) that the change in $\lambda_m^2 V$ with  $\Delta C_3$ diminishes in magnitude with the increase of $D_3$.
So $\lambda_{mt}^2 V$ will be particularly sensitive to the addition of element $3$ if the element 3 is a slow diffuser, as can be observed in fig. \ref{fig:symetrical} for two different sets of slopes for element 3. In the same way, the evolution of $\lambda_m^2 V$ with $\Delta C_3$ is inversely proportional to $\Delta m_3$. So the more similar $m^\beta_3$ and $m^\alpha_3$, the more $\lambda_m^2 V$ will change with the addition of element 3. This effect can be observed by comparing changes plotted in fig. \ref{fig:symetrical} for two different values of $\mid \Delta m_3 \mid$. 
Finally, we note from Eq. (\ref{l2Vsym}) that whether $\lambda_m^2 V$ increases or decreases with the addition of a third alloying element depends on the sign of $\Delta C_3$, $m^\beta_3$ and $m^\alpha_3$ and on their relative values. In particular, if $m^\beta_3$ and $m^\alpha_3$ have the same sign, the variation of $\lambda_m^2 V$ with $\Delta C_3$ will depend on the sign of $(m^\beta_3-m^\alpha_3)(C^\beta_3-C^\alpha_3)$.

\section{Discussion}

The eutectic growth model developed in this paper is equivalent to the Jackson Hunt theory for binary alloys. It is now admitted that the Jackson Hunt theory is satisfactory to  model eutectic growth of regular binary eutectics. 
This theory can therefore be used to analyze regular eutectics containing any number of elements and their growth properties in a similar way as the Jackson-Hunt model.
However, it has been shown for binary systems that rather than growing only with the eutectic spacing $\lambda_m$ predicted by the Jackson-Hunt theory, eutectics can grow with a range of eutectic spacing around $\lambda_m$ at a given velocity. 
Indeed, Karma and Sarkissian \cite{Karma1996} have revealed that the regular microstructure drawn on fig. \ref{fig:intro} is stable up to a critical spacing which can be as high as $2\lambda_m$. 
Akamatsu et al. \cite{Akamatsu2004} have shown experimentally and theoretically that the lower stability bound of this range of eutectic spacings can be as low as $0.7\lambda_m$. They have also observed that the eutectic spacing developed is dependent on the history of the solidification process.
So even if all parameters involved in the theory are known perfectly, the theory will only enable to give an approximate value of the eutectic spacing developed experimentally for a given velocity.\\
However, the model presented will provide guidance on how the eutectic spacing in an alloy changes with the addition of a new element through an evolution of the $\lambda_m^2 V$ law for the multicomponent system. Such an evaluation would necessitate computing thermodynamic, diffusion and curvature parameters involved in $\lambda_m^2 V$ result given above.
Among these parameters, thermodynamic coefficients (liquidus slopes and distribution coefficients) can be obtained as soon as the expression of the Gibbs free energies of the solid and liquid phases are known. Such expressions are generally gathered in thermodynamic databases such as Pandat \cite{Pandat} or ThermoCalc \cite{ThermoCalc}.
Nowadays, the thermodynamic properties of more than $10\%$ of all possible binary combinations of elements have been assessed. For ternary and quaternary  systems, thermodynamic informations are generally available for alloys presenting an industrial interest (Fe-based,Ni-based,Al-based alloys) \cite{Hecht2004}, but we are still far from knowing the thermodynamic properties of any multicomponent alloy. Nevertheless, the development of computational tools offers promising ways to accelerate our knowledge on thermodynamic properties of multicomponent systems \cite{Kattner1997}. 
Experimental values of diffusion coefficients in liquids with more than 2 elements are rare \cite{Hecht2004}. For metals, this lack of experimental studies is partly due to the fact that diffusion coefficients are particularly sensitive to fluid flow \cite{Lee2004}.
For binary and ternary mixtures, some methods are nevertheless available to compute interdiffusion coefficients from ab initio Molecular Dynamics simulations \cite{Liu2012}.
Finally, solid/liquid surface energies appear in the expression of Gibbs-Thomson coefficients and angles of curvature at the trijunction. Angles of curvature depend also on the different interphase surface energies and degree of anisotropy \cite{Magnin1991}. A review of the current knowledge on interface properties in multicomponent systems has been published by Hecht et al. \cite{Hecht2004}. They find that very little is known about surface properties in multicomponent systems, especially with more than two components. However, some experimental and numerical methods are available to obtain informations on surface properties evolution with the addition of an element, at least in dilute ternary alloys. 
Therefore, determining the evolution of surface properties with the addition of an element seems to be the most difficult part of this predictive use of the model. Computations and experiments that give these interface properties as a function of alloy composition would be very helpful. For now, we can nevertheless consider that solid/liquid surface energies are expected to decrease with the absorption of a third element \cite{Lupis1983} which would lower Gibbs-Thomson coefficients.

\section{Conclusion}

We have presented in this paper a general theory to express the mean undercooling of a two-phase eutectic containing N elements assuming that the solid/liquid interface is isothermal. 
This theory has been based on a development of the thermodynamic equilibrium at the solid/liquid interface. The expression of thermodynamic coefficients involved in the theory according to phases Gibbs free energies is presented in this paper.
It was established that the definition of a scaling parameter $\lambda_m $ such that $\lambda_m^2V=\mathrm{Constant}$ determined for binary alloys by Jackson and Hunt \cite{Jackson1966} can be generalized to alloys with N elements.

This general theory was used to establish a new model for ternary alloys two-phase eutectic growth.
It was shown that this new theory contains less approximations than previous studies on two-phase eutectics with more than two elements \cite{McCartney1980,DeWilde2004,Catalina2015}.\\
This work could be continued by developing the theory for 3D rod-like microstructures in a similar way as in the Jackson-Hunt theory \cite{Jackson1966}. Moreover, it was assumed in the theory presented that all phases have the same density, which is not the case in most alloys. It would thus be important to add the effect of these differences of density in the theory in the future. Finally, this theory has been developed by approaching the growth equations at first order in Peclet numbers. Nevertheless, this approximation may be removed by using an algorithm similar to the one of Ludwig et al. \cite{Ludwig2004} which enables to compute the growth law of the eutectic for any Peclet number value in binary alloys. 

\section*{Acknowledgments}
This work has been supported by the Dow Corning Corporation. O. S. would like to thank Silver Akamatsu and Stefan Poulsen for fruitful discussions. 

\section*{References}

\bibliography{mybibfile}

\begin{thebibliography}{10}
\expandafter\ifx\csname url\endcsname\relax
  \def\url#1{\texttt{#1}}\fi
\expandafter\ifx\csname urlprefix\endcsname\relax\def\urlprefix{URL }\fi
\expandafter\ifx\csname href\endcsname\relax
  \def\href#1#2{#2} \def\path#1{#1}\fi

\bibitem{Hillert1957}
M.~Hillert, Role of interfacial energy during solid-state phase
  transformations, Jernkontorets Annaler 141 (1957) 757--789.

\bibitem{Jackson1966}
K.~Jackson, J.~Hunt, Lamellar and rod eutectic growth, AIME Met Soc Trans 236
  (1966) 1129--1142.

\bibitem{Magnin1991}
P.~Magnin, R.~Trivedi, Eutectic growth: A modification of the jackson and hunt
  theory, Acta metallurgica et materialia 39~(4) (1991) 453--467.

\bibitem{DeWilde2004}
J.~De~Wilde, L.~Froyen, S.~Rex, Coupled two-phase [$\alpha$ (al)+ $\theta$ (al
  2 cu)] planar growth and destabilisation along the univariant eutectic
  reaction in al--cu--ag alloys, Scripta materialia 51~(6) (2004) 533--538.

\bibitem{Yamauchi1996}
I.~Yamauchi, S.~Ueyama, I.~Ohnaka, Effects of mn and co addition on morphology
  of unidirectionally solidified fesi 2 eutectic alloys, Materials Science and
  Engineering: A 208~(1) (1996) 101--107.

\bibitem{Rinaldi1972}
M.~Rinaldi, R.~Sharp, M.~Flemings, Growth of ternary composites from the melt:
  Part ii, Metallurgical Transactions 3~(12) (1972) 3139--3148.

\bibitem{Raj2001}
S.~Raj, I.~Locci, Microstructural characterization of a
  directionally-solidified ni--33 (at.\%) al--31cr--3mo eutectic alloy as a
  function of withdrawal rate, Intermetallics 9~(3) (2001) 217--227.

\bibitem{Catalina2015}
A.~Catalina, P.~Voorhees, R.~Huff, A.~Genau, A model for eutectic growth in
  multicomponent alloys, in: IOP Conference Series: Materials Science and
  Engineering, Vol.~84, IOP Publishing, 2015, p. 012085.

\bibitem{Fridberg1970}
J.~Fridberg, M.~Hillert, Ortho-pearlite in silicon steels, Acta Metallurgica
  18~(12) (1970) 1253 -- 1260.

\bibitem{McCartney1980}
D.~McCartney, J.~Hunt, R.~Jordan, The structures expected in a simple ternary
  eutectic system: Part 1. theory, Metallurgical Transactions A 11~(8) (1980)
  1243--1249.

\bibitem{DeWilde2005}
J.~De~Wilde, L.~Froyen, V.~Witusiewicz, U.~Hecht, Two-phase planar and regular
  lamellar coupled growth along the univariant eutectic reaction in ternary
  alloys: an analytical approach and application to the al--cu--ag system,
  Journal of applied physics 97~(11) (2005) 113515.

\bibitem{Ludwig2004}
A.~Ludwig, S.~Leibbrandt, Generalised ‘jackson--hunt’model for eutectic
  solidification at low and large peclet numbers and any binary eutectic phase
  diagram, Materials Science and Engineering: A 375 (2004) 540--546.

\bibitem{Donaghey1968}
L.~Donaghey, W.~Tiller, On the diffusion of solute during the eutectoid and
  eutectic transformations, part i, Materials Science and Engineering 3~(4)
  (1968) 231--239.

\bibitem{Goodman1981}
D.~A. Goodman, J.~W. Cahn, L.~H. Bennett, The centennial of the gibbs-konovalov
  rule for congruent points, Bulletin of alloy phase diagrams 2~(1) (1981)
  29--34.

\bibitem{Witusiewicz2004}
V.~Witusiewicz, U.~Hecht, S.~Fries, S.~Rex, The ag--al--cu system: part i:
  reassessment of the constituent binaries on the basis of new experimental
  data, Journal of alloys and compounds 385~(1) (2004) 133--143.

\bibitem{Witusiewicz2005}
V.~Witusiewicz, U.~Hecht, S.~Fries, S.~Rex, The ag--al--cu system: Ii. a
  thermodynamic evaluation of the ternary system, Journal of alloys and
  compounds 387~(1) (2005) 217--227.

\bibitem{Karma1996}
A.~Karma, A.~Sarkissian, Morphological instabilities of lamellar eutectics,
  Metallurgical and Materials Transactions A 27~(3) (1996) 635--656.

\bibitem{Akamatsu2004}
S.~Akamatsu, G.~Faivre, M.~Plapp, A.~Karma, Overstability of lamellar eutectic
  growth below the minimum-undercooling spacing, Metallurgical and Materials
  Transactions A 35~(6) (2004) 1815--1828.

\bibitem{Pandat}
S.-L. Chen, S.~Daniel, F.~Zhang, Y.~Chang, X.-Y. Yan, F.-Y. Xie,
  R.~Schmid-Fetzer, W.~Oates, The pandat software package and its applications,
  Calphad 26~(2) (2002) 175--188.

\bibitem{ThermoCalc}
B.~Sundman, B.~Jansson, J.-O. Andersson, The thermo-calc databank system,
  Calphad 9~(2) (1985) 153--190.

\bibitem{Hecht2004}
U.~Hecht, L.~Gr{\'a}n{\'a}sy, T.~Pusztai, B.~B{\"o}ttger, M.~Apel,
  V.~Witusiewicz, L.~Ratke, J.~De~Wilde, L.~Froyen, D.~Camel, et~al.,
  Multiphase solidification in multicomponent alloys, Materials Science and
  Engineering: R: Reports 46~(1) (2004) 1--49.

\bibitem{Kattner1997}
U.~R. Kattner, The thermodynamic modeling of multicomponent phase equilibria,
  JOM 49~(12) (1997) 14--19.

\bibitem{Lee2004}
J.-H. Lee, S.~Liu, H.~Miyahara, R.~Trivedi, Diffusion-coefficient measurements
  in liquid metallic alloys, Metallurgical and materials transactions B 35~(5)
  (2004) 909--917.

\bibitem{Liu2012}
X.~Liu, A.~Mart{\'\i}n-Calvo, E.~McGarrity, S.~K. Schnell, S.~Calero, J.-M.
  Simon, D.~Bedeaux, S.~Kjelstrup, A.~Bardow, T.~J. Vlugt, Fick diffusion
  coefficients in ternary liquid systems from equilibrium molecular dynamics
  simulations, Industrial \& Engineering Chemistry Research 51~(30) (2012)
  10247--10258.

\bibitem{Lupis1983}
C.~H. Lupis, Chemical thermodynamics of materials, Elsevier Science Publishing
  Co., Inc., 1983, (1983) 581.

\bibitem{Kurzfundamentals}
W.~Kurz, D.~Fisher, Fundamentals of solidification, 1986, Trans Tech
  Publications, Switzerland.

\end{thebibliography}

\begin{appendices}
\section{Equilibrium at the interface \label{Appendix A}}
\subsection*{Linearisation of equations}
We analyze here the thermodynamic equilibrium between the solid phase $\phi$ and the liquid phase $l$ at the interface. We suppose that this interface is curved. We note $T$ the temperature of the interface at this position, $(C^l_2,..,C^l_N)$ the composition of the liquid phase at the interface (resp $(C^\phi_2,..,C^\phi_N)$ in the solid phase $\phi$), and $p^l$ the internal pressure in the liquid phase (resp $p^\phi$).
The interface thermodynamic equilibrium implies that for every element $i=1..N$, the chemical potential of the phase $\phi$ ($\mu^\phi_i$) and of the liquid phase ($\mu^l_i$) are equal:
\begin{equation}
\mu^\phi_i(C^\phi_2,..,C^\phi_N,T,p^\phi)=\mu^l_i(C^l_2,..,C^l_N,T,p^l)
\label{slequilibA}
\end{equation}

If the temperature $T$ of the interface is close to the equilibrium eutectic temperature $T_E$, the equality (\ref{slequilibA}) can be linearly expanded about  the equilibrium state of a flat interface at the eutectic temperature:
\begin{multline}
\displaystyle \sum^N_{j=2}\frac{\partial \mu^\phi_i}{\partial C^\phi_j}\bigg|_{C^{\phi E}_k \neq C^{\phi E}_j,T_E}\Delta C^\phi_j+\frac{\partial \mu^\phi_i}{\partial T}\bigg|_{C^{\phi E}_j}\Delta T+\frac{\partial \mu^\phi_i}{\partial p}\bigg|_{C^{\phi E}_j,T_E}\Delta p^\phi=\\
\displaystyle \sum^N_{j=2}\frac{\partial \mu^l_i}{\partial C^l_j}\bigg|_{C^{l E}_j \neq C^{lE}_i,T_E}\Delta C^l_j+\frac{\partial \mu^l_i}{\partial T}\bigg|_{C^{lE}_j}\Delta T
\label{sldvlp}
\end{multline}
where all $\Delta X$ quantities correspond to the difference between the value of $X$ at the eutectic temperature and the value of $X$ at $T$: $\Delta X=X^E-X$.
In this development, we supposed that the pressure of the liquid does not change from the equilibrium state.

For the following we use the notation: $\Delta S^\phi_i=\frac{\partial \mu^\phi_i}{\partial T}\bigg|_{C^{\phi E}_j}-\frac{\partial \mu^l_i}{\partial T}\bigg|_{C^{lE}_j}$ and $\mu^\phi_{ij}=\frac{\partial \mu^\phi_i}{\partial C^\phi_j}\bigg|_{C^{\phi E}_k \neq C^{\phi E}_j,T_E}$ (we use the same notation for the liquid phase). Also $\frac{\partial \mu^\phi_i}{\partial p}\bigg|_{C^{\phi E}_j,T}=V^\phi_{m,i}$ where $V^\phi_{m,i}$ is the partial molar volume of element $i$ in the solid phase $\phi$. Using these notations, we can transform the system of N equations (\ref{sldvlp}) to the following matrix system:
\begin{equation}
\begin{bmatrix}
\mu^\phi_{12}&\dots&\mu^\phi_{1N}&\Delta S^\phi_1\\
&&\vdots\\
\mu^\phi_{N2}&\dots&\mu^\phi_{NN}&\Delta S^\phi_N
\end{bmatrix}
\times
\begin{bmatrix} 
\Delta C^\phi_2\\
\vdots\\
\Delta C^\phi_N\\
\Delta T
\end{bmatrix}
=
\begin{bmatrix}
\mu^l_{12}&\dots&\mu^l_{1N}&V^\phi_{m,1}\\
&&\vdots\\
\mu^l_{N2}&\dots&\mu^l_{NN}&V^\phi_{m,N} 
\end{bmatrix}
\times
\begin{bmatrix}
\Delta C^l_2\\
\vdots\\
\Delta C^l_N\\
-\Delta p^\phi
\end{bmatrix}
\label{eqMatrix}
\end{equation}
Defining $[A]$ as the ($N\times N$) matrix on left hand side of the matrix equation (\ref{eqMatrix}), the multiplication of this equality by the inverse of matrix $[A]$ gives a matrix equation expressing variations of concentration in the solid phase $\Delta C^\phi_i$ and the variation of temperature $\Delta T$ according to variations of concentration in the liquid phase $\Delta C^l_j$ and the variation of pressure in the solid phase $\Delta p^\phi$:
\begin{equation}
\begin{bmatrix} 
\Delta C^\phi_2\\
\vdots\\
\Delta C^\phi_N\\
\Delta T
\end{bmatrix}
=
\begin{bmatrix}
\Lambda^\phi_{22}&\dots&\Lambda^\phi_{2N}&\Theta^\phi_2\\
& \ddots&\vdots\\
\Lambda^\phi_{N2}&\dots&\Lambda^\phi_{NN}&\Theta^\phi_N\\
m^\phi_2&\dots&m^\phi_N&\Omega^\phi
\end{bmatrix}
\times
\begin{bmatrix}
\Delta C^l_2\\
\vdots\\
\Delta C^l_N\\
-\Delta p^\phi
\end{bmatrix}
\label{Linearmatrix}
\end{equation}
In this ($N\times N$) matrix, coefficients $ \Lambda^\phi_{ij}$ are called distribution coefficients and $m^\phi_i$ coefficients are the slopes of the phase $\phi$ liquidus surface corresponding to variations of concentration of elements $i$.
Coefficients of this matrix depend on partial derivatives of chemical potentials $\mu^\psi_i$ (where $\psi$ can be the solid phase $\phi$ or the liquid phase) according to independent elements concentration and temperature. These derivatives can be computed from the expressions of phases molar Gibbs free energies $G^\psi_m$ as for each element $i=1\dots N$ \cite{Lupis1983}:
\begin{equation}
\mu^\psi_i=G^\psi_m+\displaystyle \sum^N_{j=2}(\delta_{ij}-C^\psi_j)\frac{\partial G^\psi_m}{\partial C_j}\bigg|_{C^{\psi}_k \neq C^{\psi}_j,T}
\label{chemcoeff}
\end{equation}
where $G^\psi_m$ depends on independent elements concentrations $(C^\psi_2,..,C^\psi_N)$  and on temperature.

\subsection*{Curvature parameters}
In this section we analyze terms linking the variations of temperature $\Delta T$ and of elements concentration in the solid phase $\Delta C^\phi_i$ to the variation of pressure induced by the interface curvature. 

In eq. (\ref{Linearmatrix}), the coefficient $\Omega^\phi$ is defined as:
\begin{equation}
\Omega^\phi=\displaystyle \sum^N_{k=1} A^{-1}_{Nk} V^\phi_{m,k}
\end{equation}
where coefficients $A^{-1}_{ij}$ are coefficients of the inverse matrix of $[A]$. By definition $A^{-1}_{Nk}$ coefficients can be written:
\begin{equation}
A^{-1}_{Nk}=\frac{1}{\det(A)}(-1)^{N+k}B_{kN}
\end{equation}
where coefficient $B_{ij}$ is the determinant of the $(N-1)\times (N-1)$ matrix corresponding to 
$[A]$ withour row $i$ and column $j$. We note that $B_{kN}$ can be written: $B_{kN}=\det([C_k][G^\phi_{cc}])$ where $G^\phi_{cc}$ is the Hessian of phase $\phi$ Gibbs free energy (according to independent concentrations $(C^\phi_2,..,C^\phi_N)$) and $[C_k]$ is a matrix which only depends on independent elements concentration and such that $\det(C_k)=C^\phi_k(-1)^{k+1}$. In addition, $\det(A)=\displaystyle \sum^N_{k=1} (-1)^{N+k}B_{kN} \Delta S_k$. We thus get that:
\begin{equation}
\Omega^\phi=\frac{V^\phi_m}{\Delta S_{\phi l}}
\end{equation}
where $V^\phi_m=\displaystyle \sum^N_{k=1} C^\phi_k V^\phi_{m,k}$ is the molar volume of phase $\phi$ and $\Delta S_{\phi l}=\displaystyle \sum^N_{k=1} C^\phi_k \Delta S^\phi_k$ is the molar entropy of fusion of an infinitesimal amount of phase $\phi$ in the liquid phase \cite{Goodman1981}.

For $k=1\dots(N-1)$, coefficient $\Theta_{k+1}$ introduced in eq. (\ref{Linearmatrix}) is defined as:
\begin{equation}
\Theta_{k+1}=\displaystyle \sum^N_{j=1} A^{-1}_{kj} V^\phi_{m,j}
\end{equation}
which can be re-written $\Theta_{k+1}=\frac{\displaystyle \sum^N_{j=1} (-1)^{k+j}B_{jk} V^\phi_{m,j}}{\displaystyle \sum^N_{j=1} A_{jk}(-1)^{k+j} B_{jk}}$. If we suppose that terms of the same type ($A_{jk}$, $B_{jk}$, $ V^\phi_{m,k}$) have the same range of order, we obtain that $\Theta_{k+1} \sim \frac{V^\phi_{m,j}}{A_{jk}}=\frac{V^\phi_{m,j}}{\mu^\phi_{j(k+1)}}$. 
If, in addition, we assume that all terms of $[G_{cc}]$ and all elements concentration have respectively the same range of order we obtain that: 
\begin{equation}
\Theta_{k+1} \sim \frac{V^\phi_m}{\frac{\partial^2 G^\phi_m}{\partial C_{k+1} \partial C_i}}
\end{equation}
The pressure variation induced by the interface curvature is defined as: $\Delta p^\phi=\sigma_{\phi l} \kappa (x)$, where $\sigma_{\phi l}$ is the solid phase $\phi$/liquid interface energy and $\kappa (x)$ is the interface curvature in position $x$. Therefore, the effect of curvature variation on $\Delta C^\phi_i$ (for $i=2\dots N$) is of the same range of order as $\frac{V^\phi_m}{\frac{\partial^2 G^\phi_m}{\partial C_i \partial C_j}}\sigma_{\phi l} \kappa (x)$.

From the data given in Kurz and Fisher \cite{Kurzfundamentals} of pure materials, we find  $V^\phi_m \sim 10^{-5}\,m^3/\mathrm{mol}$ and  $\sigma_{\phi l} \sim 10^{-2}-10^{-1}\,J/m^2$. By only taking into account the entropy of mixing term of solid phase $\phi$ solidifying at $T \sim 10^2\,K$ we get that $\frac{\partial^2 G^\phi_m}{\partial C_i \partial C_j} \sim 10^4\,K/\mathrm{mol}$. For alloys with a eutectic spacing around $\lambda \sim 10^{-6}\,m$ we have $\kappa (x) \sim 10^6\,m^{-1}$. Based on these ranges of order, we find that the effect of curvature alone on $\Delta C^\phi_i$ (for $i=2\dots N$) is of the order of $10^{-17}-10^{-16}$, so this effect is negligible.

\subsection*{Coefficients of ternary alloys}
in this section, we develop the expression of $\Lambda^\phi_{ij}$ and $m^\phi_i$ coefficients for ternary alloys. In this particular case, the ($N\times N$) matrix defined in Eq. (\ref{Linearmatrix}) can be expressed as:
\begin{equation}
\begin{bmatrix}
\Lambda^\phi_{22}&\Lambda^\phi_{23}&\Theta^\phi_2\\
\Lambda^\phi_{32}&\Lambda^\phi_{33}&\Theta^\phi_N\\
m^\phi_2&m^\phi_3&\Omega^\phi
\end{bmatrix}
=[A]^{-1}\times 
\begin{bmatrix}
\mu^l_{12}&\mu^l_{13}&V^\phi_{m,1}\\
\mu^l_{22}&\mu^l_{23}&V^\phi_{m,2}\\
\mu^l_{32}&\mu^l_{33}&V^\phi_{m,3} 
\end{bmatrix}
\label{Linearmatrix3}
\end{equation}
where $[A]^{-1}$ is the inverse matrix of $[A]$ defined in Eq. (\ref{eqMatrix}). For ternary alloys, $[A]^{-1}$  can be expressed as:

\begin{equation}
[A]^{-1}=\frac{1}{\det(A)}
\begin{bmatrix}
B_{11}&-B_{21}&B_{31}\\
-B_{12}&B_{22}&-B_{32}\\
B_{13}&-B_{23}&B_{33}
\end{bmatrix}
\end{equation}
where\\
\begin{tabular}{ c c c}
$B_{11}=\begin{vmatrix}
\mu^\phi_{2,3}& \Delta S^\phi_2\\
\mu^\phi_{3,3}& \Delta S^\phi_3\\
\end{vmatrix}$ &
$B_{21}=\begin{vmatrix}
\mu^\phi_{1,3}& \Delta S^\phi_1\\
\mu^\phi_{3,3}& \Delta S^\phi_3\\
\end{vmatrix}$ &
$B_{31}=\begin{vmatrix}
\mu^\phi_{1,3}& \Delta S^\phi_1\\
\mu^\phi_{2,3}& \Delta S^\phi_2\\
\end{vmatrix}$\\
\addlinespace[2ex]
$B_{12}=\begin{vmatrix}
\mu^\phi_{22}& \Delta S^\phi_2\\
\mu^\phi_{32}& \Delta S^\phi_3\\
\end{vmatrix}$ & 
$B_{22}=\begin{vmatrix}
\mu^\phi_{12}& \Delta S^\phi_1\\
\mu^\phi_{32}& \Delta S^\phi_3\\
\end{vmatrix}$ &
$B_{32}=\begin{vmatrix}
\mu^\phi_{12}& \Delta S^\phi_1\\
\mu^\phi_{22}& \Delta S^\phi_2\\
\end{vmatrix}$\\
\addlinespace[2ex]
$B_{13}=\begin{vmatrix}
\mu^\phi_{22}&\mu^\phi_{23}\\
\mu^\phi_{32}&\mu^\phi_{33}\\
\end{vmatrix}$ &
$B_{23}=\begin{vmatrix}
\mu^\phi_{33}&\mu^\phi_{32}\\
\mu^\phi_{13}&\mu^\phi_{12}\\
\end{vmatrix}$ &
$B_{33}=\begin{vmatrix}
\mu^\phi_{12}&\mu^\phi_{13}\\
\mu^\phi_{22}&\mu^\phi_{23}\\
\end{vmatrix}$
\end{tabular}
\\We thus obtain that 
\begin{eqnarray}
\Lambda^\phi_{ij}=\frac{\displaystyle \sum^3_{k=1}(-1)^{k+i-1}B_{k(i-1)}\mu^l_{kj}}{\displaystyle \sum^3_{k=1}(-1)^{k+i-1}B_{k(i-1)}\mu^\phi_{ki} }\\
m^\phi_i=\frac{\displaystyle \sum^3_{k=1}(-1)^{k+3}B_{k3}\mu^l_{ki}}{\displaystyle \sum^3_{k=1} (-1)^{k+3}B_{k3}\Delta S^\phi_k}
\end{eqnarray}

\section{Approximations of the model \label{Appendix B}}

\subsection*{Approximation of $\Delta'$ as independent of growth conditions}
In this appendex, we analyze the hypotheses that $-\Delta'=\frac{\partial \overline{\Delta T}^\beta}{\partial f_\alpha}\bigg|_{f^E_\alpha}- \frac{\partial \overline{\Delta T}^\alpha}{\partial f_\alpha}\bigg|_{f^E_\alpha}$ can be approximated to $-\Delta'_0$ where $\Delta'_0$ is independent of $\lambda$ and $V$.
The expression of $\Delta'$ is obtained from expressions of $\overline{\Delta T}^\alpha$ (Eq. (\ref{dTalpha})) and $\overline{\Delta T}^\beta$ (Eq. (\ref{dTbeta})) and from mean liquid concentration of independent elements on each solid phase interface (Eq. (\ref{Cilalpha}) and (\ref{Cilbeta})).
From these expressions, the derivation of $\overline{\Delta T}^\alpha$ and $\overline{\Delta T}^\beta$ according to $f_\alpha$ induce that $\Delta'$ can be written:
\begin{equation}
\Delta'=\Delta'_0+V \lambda \xi_C+\frac{\xi_R}{\lambda}
\label{Deltapexp}
\end{equation}
where $\Delta'_0$, $\xi_C$ and $\xi_R$ are coefficients independent on $\lambda$ and $V$.\\
As $\Delta'_0$ is a zero order term in Peclet numbers and $ V \lambda \xi_C$ is at first order term in Peclet numbers, if the interface grows at low Peclet numbers, then we can assume that $V \lambda \xi_C \ll \Delta'_0$. Moreover, by analyzing the expressions of $\overline{\Delta T}^\beta$ and $ \overline{\Delta T}^\alpha$ we find:
\begin{equation}
\Delta'_0=\displaystyle\sum_{i=2}^N \Delta m_i \displaystyle\sum_{j=2}^N \left[ \overline{\Lambda}\right]^{-1}_{ij} \Delta C_j 
\end{equation}
In the general case, the range of order of parameters involved in this expression are: $\Delta m_i\sim (10-10^3)\,K/(\mathrm{at}\%)$, $\Delta C_j \sim 10\,\mathrm{at}\%$ and $\left[ \Lambda\right]^{-1}_{ij} \sim 0.1-10$, so $\Delta'_0 \sim 10-10^5 \,K$. We also get that 
\begin{equation}
\frac{\xi_R}{\lambda}=2\left[ \frac{\Gamma_{\beta /l}\sin(\mid \theta_\beta \mid)}{\lambda f^2_\beta}+\frac{\Gamma_{\alpha /l}\sin(\mid \theta_\alpha \mid)}{f^2_\alpha} \right] 
\end{equation}
The range of order of parameters involved in this expression are: $\Gamma_{(\alpha/\beta) /l} \sim 10^{-7}\,K.m$, $\sin(\mid \theta_{(\alpha /\beta)} \mid)\sim 10^{-1}$, $f_{(\alpha/\beta)} \sim 10^{-1}$ and $\lambda \sim 10^{-6}\,m$. So $\frac{\xi_R}{\lambda} \sim 1\,K$ and so, in the general case, $\frac{\xi_R}{\lambda} \ll \Delta'_0$.
 
We verify now that $\left(\xi_C V \lambda_m, \xi_R/\lambda_m \right) \ll \Delta'_0$ (where $\lambda_m$ is the eutectic spacing of minimum undercooling for a given velocity) on 4 binary alloys: $\mathrm{Fe-Fe_3C}$, $\mathrm{Al-Si}$, $\mathrm{Al-Al_2Cu}$ and $\mathrm{Sn-Pb}$. Parameters used for this study are taken from Ref \cite{Magnin1991}.
For $V=100\times 10^{-6}\,\mathrm{m/s}$ we obtain that for all systems, $V \lambda_m \xi_C$ and $\frac{\xi_R}{\lambda_m}$ are three orders of magnitude smaller than $\Delta'_0$. So the approximation $\Delta' \approx \Delta'_0$ is relevant for all these systems.

\subsection*{Is $\Delta f^{iso}_\alpha$ a first order term in $Pe_i$?}
We supposed in section \ref{sec:meanUndercooling} that the term $\frac{\partial \overline{\Delta T}_C}{\partial f_\alpha}\bigg|_{f^E_\alpha} $ could be approximated at zero order in Peclet number in the expression of the mean undercooling of an isothermal interface (see Eq. \ref{dTiso1}). This assumption is justified if $\Delta f^{iso}_\alpha$ is a term at first order in Peclet numbers and so if the $\alpha_R/\lambda_m$ term in Eq (\ref{dfnumerator}) is of the same range of order as $\alpha_C V\lambda_m$. We analyze this hypotheses on the 4 binary systems used in the first part of appendix \ref{Appendix B}.
We observe that $\alpha_R/\lambda_m$ is of the same range as $\alpha_C V \lambda_m$ for all systems except for $\mathrm{Sn-Pb}$ where $\alpha_C V \lambda_m=-0.17\frac{\alpha_R}{\lambda_m}$.
This induces that, for $\mathrm{Sn-Pb}$, $\Delta f^\mathrm{iso}_\alpha$ is a variation at a lower order than $Pe_i$ and that, for this system, $\frac{\partial \overline{\Delta T}_C}{\partial f_\alpha}\bigg|_{f^E_\alpha} $ could be developed at first order in Peclet number in eq. (\ref{dTmeanIso}).  
\end{appendices}

\end{document}